\documentclass[12pt,preprint]{aastex}

\shorttitle{X-ray sources in NGC~720}

\shortauthors{Jeltema et al.}

\begin{document}

\title{X-ray Source Population in the Elliptical Galaxy NGC~720 with Chandra}

\author{Tesla E. Jeltema and Claude R. Canizares}

\affil{Center for Space Research and Department of Physics}
\affil{Massachusetts Institute of Technology}
\affil{Room 37-241, 70 Vassar Street, Cambridge, MA 02139-4307}

\email{tesla@space.mit.edu, crc@space.mit.edu}

\author{David A. Buote}

\affil{Department of Physics and Astronomy}
\affil{University of California at Irvine}
\affil{4129 Frederick Reines Hall, Irvine, CA 92697-4575}

\email{buote@uci.edu}

\author{Gordon P. Garmire}
\affil{The Pennsylvania State University}
\affil{525 Davey Lab, University Park, PA 16802}

\email{gpg2@psu.edu}

\begin{abstract}

With a {\itshape Chandra\/} ACIS-S3 observation, we detect 42 X-ray point sources in the elliptical galaxy NGC~720, including a possible central source.  Most of these sources will be low-mass X-ray binaries (LMXBs), and 12 are located within 2$^{\prime\prime}$ of globular cluster candidates.  We investigate both the hardness ratios and combined spectra of the sources.  They exhibit a distribution of X-ray colors similar to those seen in other early-type galaxies.  We find that there is a population of highly absorbed sources located at large distances from the center of the galaxy.  The overall spatial distribution of sources is consistent with the ellipticity and position angle of the galaxy, but the sources appear to form several arcs.  NGC~720 contains nine ultraluminous sources (L$_{x} \geq 10^{39}$ ergs s$^{-1}$).  This number is more than have previously been detected in an early-type galaxy but similar to the number seen in the Antennae merger system.  The ratio $L_{ULX}/L_B$ for NGC~720 is more than double the ratio for the S0 galaxy NGC~1553 and a factor of seven higher than for the elliptical galaxy NGC~4697, although uncertainties in the distance and the source spectral properties could bring these ratios into agreement.  The X-ray source luminosity function is also nearly as flat as those seen in disk and merger systems.  The large number of ULXs in NGC~720 and its relatively flat XLF may provide evidence against the association of all ULXs with young stars.  We also see a possible high luminosity break in the luminosity function at $2 \times 10^{39}h_{50}^{-2}$ ergs s$^{-1}$.

\end{abstract}


\section{INTRODUCTION}

Early X-ray observations of elliptical galaxies revealed that the dominant source of X-ray emission comes from the diffuse, hot interstellar medium (ISM) at a temperature of $\sim$1 keV (e.g., Forman, Jones, \& Tucker 1985).  However, these galaxies have also been shown to contain a population of X-ray emitting point sources (Sarazin, Irwin, \& Bregman 2000; Fabbiano, Kim, Trinchieri 1994).  Before the launch of {\itshape Chandra\/}, with its excellent angular resolution, X-ray source populations could not be studied in detail in galaxies outside the Local Group.  However, many studies were done on the estimated cumulative emission from discrete sources with previous X-ray satellites (Kim, Fabbiano, \& Trinchieri 1992; Matsumoto et al. 1997; Irwin \& Sarazin 1998).  The presence of these sources could be seen as deviations from the spectral and spatial properties of the hot gas.  In NGC~720, for example, the total emission from discrete sources has been estimated from both the hard spectral component detected with {\itshape ASCA\/} and from the spatial distribution of the X-ray surface brightness with {\itshape ROSAT\/}.  In both cases, the ratio of the flux of discrete sources to the flux of diffuse emission was estimated to be about 0.40 (Buote \& Canizares 1996, 1997).  With {\itshape Chandra\/}, the sources in a number of galaxies of varying morphological type are now being detected and compared.  Early-type galaxies studied include the faint elliptical and S0 galaxies NGC~4697 and NGC~1553 (Sarazin et al. 2000; Blanton, Sarazin, \& Irwin 2001) as well as the cluster elliptical NGC~1399 (Angelini, Loewenstein, \& Mushotzky 2001).  The X-ray Luminosity Functions (XLF) of sources in these galaxies have been found to follow a broken power-law distribution with a break luminosity close to the Eddington limit for an accreting neutron star, and they contain a number of sources with luminosities exceeding this limit.  Among late-type galaxies studied, Kilgard et al. (2002) compare the luminosity functions of X-ray sources in a few spiral and starburst galaxies, Zezas et al. (2002) presents a detailed study of the sources in the Antennae galaxies, and Lira et al. (2002) looks at the sources in the luminous IR galaxy NGC~3256.

It was found early on that some galaxies contain ultraluminous sources (L$_{x} \geq 10^{39}$ ergs s$^{-1}$, Fabbiano 1989).  The nature of these sources has not yet been determined.  The possibilities include intermediate mass black holes, with masses between those of stellar black holes and AGN (e.g. Colbert \& Mushotzky 1999), or beamed emission from stellar mass black hole or neutron star binaries (King et al. 2001; King 2002).  {\itshape Chandra\/} has revealed these sources in many galaxies.  Comparison of the XLFs of galaxies of different morphological type has led to the suggestion that a larger fraction of ultraluminous sources is present in star forming galaxies and that they may be associated with young stellar populations (Kilgard et al. 2002; Zezas \& Fabbiano 2002).  

In this paper, we study the previously unresolved X-ray point source population in NGC~720.  NGC~720 is an X-ray bright, relatively isolated elliptical galaxy.  An AGN has not previously been detected in this galaxy, although as discussed later we see possible evidence for a low-luminosity AGN with {\itshape Chandra\/}.  Some of the galaxy's basic properties are listed in Table 1.  The diffuse X-ray emission in NGC~720 was studied previously with {\itshape ROSAT\/} and {\itshape ASCA\/} (Buote \& Canizares 1994, 1996, 1997) and is studied in detail with {\itshape Chandra\/} in a separate paper (Buote et al. 2002).  The diffuse emission shows a position angle twist consistent with a triaxial mass distribution.  This isophote twist is not seen in the optical, and the ellipticity of the X-ray emission is too large to be explained by a model with mass following optical light.  Therefore, the distribution of X-ray emission indicates the presence of dark matter in this galaxy.  

The distance to NGC~720 has been determined through a variety of methods.  The distance measured from surface brightness fluctuations (SBF) calibrated to Cepheids is $28\pm5$ Mpc (Tonry et al. 2001), and the fundamental plane (FP) distance estimate is $35^{+7}_{-6} h_{50}^{-1}$ Mpc (Blakeslee et al. 2001; Blakeslee et al. 2002).  Blakeslee et al. (2002) find that combination of the SBF and FP distances for a large sample of galaxies gives $H_0 = 68$ km s$^{-1}$ Mpc$^{-1}$, and for this value of $H_0$ the FP and SBF distances to NGC~720 agree well.  However, we use $H_0 = 50$ km s$^{-1}$ Mpc$^{-1}$ in order to compare to the results of Sarazin et al. (2000) and Blanton et al. (2001).  We, therefore, adopt a distance to NGC~720 of 35$h_{50}^{-1}$ Mpc; at this distance 1$^{\prime\prime} \sim 0.17 kpc$.  For $H_0 = 68$ km s$^{-1}$ Mpc$^{-1}$, all our luminosities should be reduced by a factor of 1.8.

\placetable{tbl-1}

\section{OBSERVATION AND DATA REDUCTION}

NGC~720 was observed with the back-illuminated {\itshape Chandra\/} ACIS-S3 detector on 2000 October 12-13 for 41,565 s.  We kept only {\itshape ASCA\/} grades 0, 2, 3, 4, and 6\footnote{{\itshape Chandra\/} Proposers' Observatory Guide http://asc.harvard.edu/udocs/docs/docs.html, section ``Technical Description'', ``ACIS''}.  We then examined the satellite aspect and light curve to eliminate time intervals of bad aspect or high background.  Columns adjacent to node boundaries were also removed.  After filtering, the net useful exposure time was 38,830 s.  The 0.3-10 keV image of NGC~720 is shown in Figure 1.  The image has been smoothed with the Chandra Interactive Analysis of Observations (CIAO) program {\itshape csmooth\/} with a minimum significance of 3 and a maximum significance of 5.  The minimum, maximum, and average smoothing scales were 0.188, 74.7, and 62.6 pixels respectively.  Also shown is the X-ray image overlaid with the optical contours from the DSS image of NGC~720.  After our analysis of NGC~720 was complete, new alignment files were released which correct for aspect offsets in {\itshape Chandra\/} observations based on the cross-correlation of Chandra sources and several accurate catalogs (ICRS, Tycho2, USNO-A2.0, 2MASS).  The new alignment values are based on a decaying exponential fit to the offsets in a carefully selected sample of observations spanning the entire mission.  For a sample of observations reprocessed with the new alignments files the 90\% uncertainty in the {\itshape Chandra\/} positions for sources within 2$^{\prime}$ of the HRMA optical axis was 0.6$^{\prime\prime}$\footnote{http://asc.harvard.edu/cal/ASPECT/celmon/index.html}.  The offset from the original positions for NGC~720 was $\delta$R.A. = -0.93$^{\prime\prime}$ and $\delta$Dec. = 0.27$^{\prime\prime}$.  We have corrected all of the source positions in this paper.  We were not able to find any optical or radio sources with accurate positions that had X-ray counterparts on ACIS-S3, so we were not able to independently check the astrometry.

Before spectral fitting, the data were corrected for charge transfer inefficiency (CTI) and matching RMFs were used (Townsley 2000).  The spectral analysis was limited to the 0.3-7 keV range to reduce the effect of the background.  At the time of this paper, there was some uncertainty in the spectral calibration at low energies.  To investigate the effects of this problem, we tried both using a newer test version of the quantum efficiency file which does a better job of modeling the carbon K edge region of the spectrum and fitting only energies above 0.7 keV.  However, almost all of our fits were the same within the errors, and the luminosities changed by less than 10\%.  Only the fits in the 0.3-7 keV range are reported here.  All spectra were extracted in PI (pulse height-invariant) channels, which correct for the gain difference between different regions of the CCD, and grouped to give a minimum of 15 counts per energy bin.  Fitting was done with XSPEC (v11.1.0).  

\placefigure{fig-1}

\section{SOURCE DETECTION}

Point sources were detected using {\itshape wavdetect\/}, a wavelet source detection program in CIAO. The energy range was restricted to 0.3-10 keV, the wavelet scales used were 2, 4, 8, 16, and 32 pixels, and the significance threshold was set at $10^{-6}$, corresponding to about one false detection in the area of the S3 chip.  Sources were then examined by eye in both the smoothed image and binned raw data.  Sources within approximately $2.5^{\prime}$ of the galaxy center are listed in Table 2 in order of increasing distance from the optical center.  This list includes all sources inside the $D_{25}$ ellipse.  There are a total of 42 sources including source 1, which is a possible central source.  Table 2 lists the source number and position, the projected distance from the optical center of NGC~720, the count rate, the significance of the detection, and the unabsorbed X-ray luminosity assuming the sources are located at the distance of NGC~720.  Details of the luminosity calculation are given in section 4.3.  Source counts were calculated in circular apertures surrounding each source with radii of $1.5^{\prime\prime}-5^{\prime\prime}$ depending on source strength, crowding, and distance from the aim point, and the background was estimated from an annular local background.  The count rates given by {\itshape wavdetect\/} were not used because they were too high for a few sources near the center of the galaxy, most likely due to the diffuse emission  For the other sources, {\itshape wavdetect\/} gave very similar count rates to those listed.  Examination of the exposure map for this observation showed that the exposure varied by 3\% or less between sources, so we have used the same exposure time for all sources.  Errors in the count rate were calculated using the Gehrels approximation (Gehrels 1986), and the source significances were estimated based on the Poisson probability of detecting the observed total source counts given the expected number of counts from the background.  The minimum detected count rate was approximately $2.5 \times 10^{-4}$ counts s$^{-1}$.  Our detection limit may be a bit higher near the central region of the galaxy where the diffuse emission is brightest, but examination of Table 2 shows that we are still detecting relatively low count rate sources at high significance at a radius of approximately 20$^{\prime\prime}$.  Based on source number counts from the {\itshape Chandra\/} Deep Field South and {\itshape Chandra\/} Deep Field North, we expect that 8-11 of the sources are not associated with the galaxy (Campana et al. 2001, Tozzi et al. 2001, Brandt et al. 2001).  The fits of Mushotzky et al. (2000) and Baldi et al. (2001) give a slightly larger number of expected unassociated sources.  However, their fits require extrapolation down to our flux limit, and, from the deeper surveys, it appears that the LogN-LogS flattens at small fluxes.

\placetable{tbl-2}

We have compared our X-ray source positions to a list of globular cluster candidates provided by M. Kissler-Patig (Chapelan 2001).  These globular cluster positions were derived from observations of NGC~720 taken by Buat and Burgarella at the 3.6 meter Canada-France-Hawaii Telescope (CFHT), and globular cluster candidates were selected based on their B-I color.  We find that eight X-ray sources are a distance of 1$^{\prime\prime}$ or less from a globular cluster candidate and another four are 2$^{\prime\prime}$ or less.  The globular cluster positions have an accuracy of about 1-2$^{\prime\prime}$.  Given the density of globular clusters we would expect $\sim0.4$ random associations within 1$^{\prime\prime}$ and $\sim1.5$ within 2$^{\prime\prime}$.  For this calculation, the globular cluster density was calculated in five radial annuli about the galaxy center.  The possible globular cluster associations are noted as ``GC'' in Table 2.

\section{SOURCE PROPERTIES}

\subsection{Spectral Properties}

With the exception of the brightest source, number 40, the sources have too few counts to fit individual spectra.  However, we can get an indication of their spectral properties through hardness ratios.  Following Sarazin et al. (2000) and their study of the sources in NGC 4697, we divide source counts into three energy bands: soft S, 0.3-1.0 keV; medium M, 1.0-2.0 keV; and hard H, 2.0-10.0 keV.  We then plot H31 = (H-S)/(H+S) versus H21 = (M-S)/(M+S) for all sources with more than 20 counts.  This color-color diagram is shown in Figure 2.  Error bars give 1-sigma errors in the ratios based on a binomial distribution of the source counts in each band, but do not include error due to the background.  Also plotted are the predicted ratios for different power-law spectral models.  The upper line is for a power-law with a hydrogen column density equal to the Galactic value of $1.55 \times 10^{20}$ cm$^{-2}$ (Dickey \& Lockman 1990); the plotted points are for photon indices of 0, 1, 2, and 3.  The lower line gives the colors for a column of $3.0 \times 10^{21}$ cm$^{-2}$ and photon indices of 1, 2, 3, and 4.  Overall, the distribution of source colors is quite similar to those seen in the faint elliptical and S0 galaxies NGC~4697 and NGC~1553 (Sarazin et. al 2000; Blanton et al. 2001). 

\placefigure{fig-2}

The color-color diagram reveals that there are several sources with positive hardness ratios, indicating that they emit mostly high energy photons.  These sources appear to be absorbed.  Most of the other sources are in a diagonal band with negative H31 and H21 ranging from -0.3 to 0.2.  There is also one very soft source, number 3, with ratios of (H21, H31) = (-0.6, -0.9).  Both the brightest source, source number 40, and the central source number 1 have high hardness ratios of (0.5, 0.5) and (0.3, 0.2) respectively.  In order to look at their composite spectra, we divided the sources into three groups according to their hardness ratios.  The sources in group A were defined to be those with both H21 and H31 greater than zero.  The sources in this group are indicated with an ``A'' in the last column of Table 2 and are plotted with open diamonds in Figure 2.  Source 40 and the central source were excluded from the group A composite spectrum, but both are denoted with an ``A'' and plotted with diamonds.  Source 40 was not included in the composite spectrum because it has a significantly higher flux than the other sources and could bias the overall spectrum.  We fit its spectrum separately.  The central source was not included due to the possibility that it is an AGN, however, its inclusion would not significantly affect the fits.  The sources in group B were those sources with H31 less than zero and which fell closer to the galactic absorption curve in Figure 2.  These sources are indicated with ``B'' in Table 2 and plotted with open squares in Figure 2.  The group C sources, which lie closer to the curve for a column of $3.0 \times 10^{21}$ cm$^{-2}$, are denoted as ``C'' and plotted with Xs.

An interesting property of the group A sources is that, with the exception of source 6, they all lie more than an arcmin from the galaxy center.  Of the fourteen sources at a distance of greater than $1.5^{\prime}$, nine are A sources.  For contrast, within $1.5^{\prime}$ of the center, only five of twenty-seven sources are from group A (excluding the center).  This effect could be due to the fact that at larger radius more sources will be unassociated with the galaxy.  We could account for nearly all of the these sources with the expected number of unrelated sources.  However, four of the group A sources (including number 40) seem to be associated with globular cluster candidates, and globular clusters are seen at radii larger than those considered here (Chapelan 2001).

We fit the combined spectrum of all of the sources (excluding 1 and 40) as well as the combined spectra for each of the three groups.  The results are summarized in Table 3.  For a power-law fit, we find that both the group A and C sources are consistent with having absorption greater than the Galactic value of $1.55 \times 10^{20}$ cm$^{-2}$ (Dickey \& Lockman 1990) with the A sources having a higher best-fit column of $7 \times 10^{21}$ cm$^{-2}$.  In both cases, setting the absorption to the Galactic value gave a worse fit.  Both groups have a similar photon index of about 2.  For the group B sources, the best fit column was below Galactic, so we fixed it at the Galactic value.  These sources had a flatter index than the other two groups with a best-fit value of 1.2.  From the spectra, it appears that the group A sources are actually softer than the B sources; they appear hard from their hardness ratios because their high absorption results in a small number of low energy photons.  The combined spectrum of all of the sources, not surprisingly, gave values intermediate between these with a best-fit column density of $1.1 \times 10^{21}$ cm$^{-2}$ and a best-fit index of 1.5.  All of the above fits had a reduced $\chi^2$ of one or less.  The A, B, and C spectra and best-fit power-law models are shown in Figure 3.

\placetable{tbl-3}

\placefigure{fig-3}

For both the C and total source spectra, we found that we could obtain a good fit for Galactic absorption by adding a black body thermal component to the power-law model.  For both spectra, we then get a power-law index of $\sim1$, in good agreement with the group B index, and black body temperatures of 0.38$^{+0.07}_{-0.06}$ keV and 0.62$^{+0.15}_{-0.11}$ keV respectively.  Both the single component and power-law plus black body models gave excellent fits to the data.  For the spectrum of all of the sources, the black body component accounted for approximately 30\% of the flux, while for group C, the black body component contributed approximately 40\% of the flux.  This was not a good model for either the A or B spectra, giving unreasonable values of the parameters (kT$\sim0$) for the A sources and an essentially unconstrained black body temperature for the B sources.  

We also tried fitting to an absorbed thermal bremsstrahlung model.  The fits to the A and C sources gave temperatures of 6 keV and 3 keV, with the A sources again having a higher column.  For the composite spectrum of all the sources, we got a higher best-fit temperature of 11 keV.  We do not include a bremsstrahlung fit to the group B spectrum because the temperature had very large errors, and we conclude that it is not well described by this model.

Also shown in Table 3 are the results of fits to the spectrum of source 40.  Of the 42 sources in Table 2, only source 40 had enough counts to fit an individual spectrum.  As mentioned before, this source is hard and located a distance of 134$^{\prime\prime}$ from the center of the galaxy.  A power-law fit to the spectrum gives a column density of 4.3$^{+4.3}_{-2.5} \times 10^{21}$ cm$^{-2}$, in excess of the Galactic value, and an index of 1.3$^{+0.7}_{-0.4}$.  Therefore, this source may be an AGN unassociated with the galaxy.  If it were located at the same distance as NGC~720, it would have a luminosity of $1.1 \times 10^{40}$ ergs s$^{-1}$ which is three times as luminous as the next most luminous source.  However, source 40 is located 1.6$^{\prime\prime}$ from a globular cluster candidate.  We also fit to a black body model, which gave a temperature of 0.93$^{+0.16}_{-0.13}$ keV.  Here, we chose to fix the absorption at the Galactic value because the best fit was below this value.  The spectrum and best-fit power-law model are shown in Figure 4.

\placefigure{fig-4}

\subsection{Central Source}

As mentioned earlier, we detect a source close to the optical center of the galaxy.  It is listed as source 1 in Table 2.  The position of the center is listed in Table 1 and was taken from the NASA/IPAC Extragalactic Database (NED).  With hardness ratios of (H21, H31) = (0.3, 0.2), this source appears to be either hard or highly absorbed.  These ratios are harder than those for the diffuse galaxy component, and this source is not likely to be simply a peak in the diffuse emission.  A Kolmogoroff-Smirov test of the events in source 1 versus the diffuse emission in the central 15$^{\prime\prime}$ of NGC~720 gives a probability of 0.13\% that they were drawn from the same distribution of energies.  A KS test of source 1 versus the diffuse emission from the whole galaxy gives a probability of 1.3\%.  {\itshape Wavdetect\/} finds the size of this source to be fairly large given its position near the aim point, with a radius of over 3$^{\prime\prime}$, indicating that it could be extended.  However, this could be due to contamination from the bright diffuse emission at the galaxy center, and the source appears small in both the smoothed image and binned data.  Using the group A spectrum normalization, we find that source 1 has a luminosity of $2.5 \times 10^{39}h_{50}^{-2}$ ergs s$^{-1}$.  The high luminosity, large hardness ratios, and location of this source indicate that it could be a low-luminosity AGN (Terashima \& Wilson 2002; Ho et al. 2001).  However, we see several other ultraluminous sources in NGC~720 which have similar hardness ratios or are associated with globular cluster candidates.  An AGN has not been detected in this galaxy at other wavelengths, but from the central surface birghtness profile van der Marel (1999) estimated a  black hole mass of $2.29 \times 10^8 M_{\odot}$ for this galaxy.  A correlation has also been found between bulge luminosity and black hole mass (Kormendy \& Gebhardt 2001).  Using the blue luminosity of NGC~720 and the fit of Kormendy \& Gebhardt (2001), we estimate a black hole mass of $3 \times 10^8 M_{\odot}$.  If this source is a central AGN with this black hole mass, the X-ray luminosity is significantly below the Eddington luminosity ($\sim10^{-7} L_{edd}$).  This relative lack of nuclear activity has been seen in several other elliptical galaxies with central black holes (Loewenstein et al. 2001).

\subsection{Luminosity Function}

We calculated the source luminosities assuming they were all located at the distance of NGC~720 for $H_0 = 50$ km s$^{-1}$ Mpc$^{-1}$.  When constructing the luminosity function, we converted from count rate to unabsorbed luminosity in the 0.3-10 keV band using the best-fit power-law to the composite spectrum of all of the sources which has a photon index of 1.5 and a column density of $1.1 \times 10^{21}$ cm$^{-2}$.  The conversion factor is $1.33 \times 10^{42}$ ergs cnt$^{-1}$.  The individual luminosities listed in Table 2 were calculated using the best-fit power-laws to each group of sources, and they are absorption corrected using the best-fit absorption.  The conversions used are $2.48 \times 10^{42}$ ergs cnt$^{-1}$ for group A, $1.12 \times 10^{42}$ ergs cnt$^{-1}$ for group B, and $1.31 \times 10^{42}$ ergs cnt$^{-1}$ for group C.  The luminosity of source 40 was calculated using its individual spectrum.  The central source (source 1) count rate was converted to luminosity using the group A normalization, but it was not included in the luminosity function.  The sources all have luminosities between $4 \times 10^{38}$ ergs s$^{-1}$ and $10^{40}$ ergs s$^{-1}$, meaning that they all have luminosities exceeding the Eddington limit of $2 \times 10^{38}$ ergs s$^{-1}$ for a 1.4 $M_{\odot}$ neutron star.  The luminosity function of the sources is shown as a histogram in Figure 5.  Due to possible incompleteness at low luminosities, we have only included sources with luminosities above $5 \times 10^{38}$ ergs s$^{-1}$; this luminosity corresponds roughly to a detection significance of 4.5 $\sigma$.  We detect sources with this luminosity at high significance in to a radius of approximately 20$^{\prime\prime}$ from the galaxy center.  We are, therefore, probably not missing more than a couple of sources due to the higher level of diffuse emission in the central regions of the galaxy.

We fit the luminosity function to a broken power-law of the form 
\begin{equation}
\frac{dN}{dL_{38}} = N_0\biggl(\frac{L_{38}}{L_b}\biggr)^{-\alpha},
\end{equation}
where $\alpha = \alpha_l$ for $L_{38} \leq L_b$, $\alpha = \alpha_h$ for $L_{38} > L_b$, and $L_{38}$ is the luminosity in the 0.3-10 keV range in units of $10^{38}$ ergs s$^{-1}$ (Sarazin et al. 2000).  The luminosity function was fit using the maximum-likelihood method.  In the fit, we included a third power-law to account for sources unrelated to the galaxy based on the results of Tozzi et al. (2001) from the {\itshape Chandra\/} Deep Field South.  Unfortunately, all of the deep surveys publish LogN-LogS in two separate bands, 0.5-2 keV and 2-10 keV.  For the background term, we used their 2-10 keV LogN-LogS fit adjusting the luminosities to our 0.3-10 keV band using the best-fit power-law to the composite spectrum of all the sources.  This gives a slightly larger number of background sources than if we had used the 0.5-2 keV fit.  For $H_0 = 50$ km s$^{-1}$ Mpc$^{-1}$ and source luminosities determined from the spectrum of all of the sources, this fit gave a normalization N$_0 = 0.6\pm0.1$, upper and lower slopes $\alpha_h = 5^{+6}_{-2}$ and $\alpha_l = 1.4\pm0.3$, and a break luminosity of L$_b = 2.1\pm0.2 \times 10^{39}$ ergs s$^{-1}$.  The errors are 1$\sigma$ and are calculated from the change in the likelihood.  This model is plotted over the data in Figure 5.  The normalization and low luminosity slope are slightly anticorrelated.  The best-fit single power-law has a slope of $2.3\pm0.1$.  Based on the change in the likelihood statistic, a broken power-law gave a better fit than a single power-law by 2.1 $\sigma$.  A Kolmogoroff-Smirov test gave probabilities of 63\% for the broken power-law model and 8.7\% for the single power-law that the data are drawn from the respective model.  Because we first fit to the data, these are essentially the maximum probabilities the K-S test will give for these models.  We, therefore, see a marginal rejection of the single power-law model, but we can not reject the broken power-law.  The best-fit single power-law is also plotted in Figure 5.  From the plot, it is apparent that a single power-law will either overestimate the total number of sources or overestimate the number of high luminosity sources.  If the two most luminous sources were background objects, the XLF could be represented by a cutoff power-law.  At the luminosity of these sources, we expect about one source to be unassociated with the galaxy (Tozzi et al. 2001).  Using instead the luminosities from the best-fit spectra of each group and a minimum luminosity of $6.5 \times 10^{38}$ ergs s$^{-1}$, a broken power-law fit gave L$_b = 1.0\pm0.5 \times 10^{39}$ ergs s$^{-1}$, N$_0 = 4\pm1$, $\alpha_h = 2.8\pm0.5$, and $\alpha_l = -7\pm5$.  In the rest of the paper, we will use the fit to the XLF assuming a single source spectrum when comparing to the results for other galaxies.  In the relevant, literature a single source spectrum is assumed.

The luminosity functions of the X-ray sources in the faint early-type galaxies NGC~4697 and NGC~1553 were both well fit by broken power-laws (Sarazin et al. 2000; Blanton et al. 2001).  In these galaxies, a break in the luminosity function is found at 3-4 $\times 10^{38}$ ergs s$^{-1}$, and the high luminosity slopes are $2.76^{+1.81}_{-0.39}$ and $2.7^{+0.7}_{-0.4}$ (Sarazin et al. 2000; Blanton et al. 2001).  These slopes are flatter than our high-end broken power-law slope and somewhat steeper than our single power-law slope, but they all roughly agree within the errors.  Our minimum luminosity is too high to compare to their break luminosity, however, we see a possible break at a significantly higher luminosity.  We have used the same value of the Hubble constant, $H_0 = 50$ km s$^{-1}$ Mpc$^{-1}$, as was used in these papers.  For NGC~4697 and NGC~1553, the break luminosity is a few times the Eddington limit for a neutron star, and Sarazin et al. (2000) suggest that this break might indicate a transition from neutron star to black hole binaries.  This hypothesis would not explain the high break luminosity in NGC~720.  Alternatively, it has been suggested that a break in the XLF could be produced by a decaying starburst component from binaries formed in a past merger and star-forming episode (Kilgard et al. 2002; Wu 2001).  If this is the case, NGC~720 may have undergone a more recent merger than either NGC~4697 or NGC~1553.  As discussed later in Section 4.3, the non-uniform distribution of X-ray sources in NGC~720 may be an indication of a past merger.

NGC~720 has more ultraluminous (ULXs, L$_{x} \geq 10^{39}$ ergs s$^{-1}$) sources than has previously been seen in an early-type galaxy.  For a commonly accepted value of the Hubble constant, $H_0 = 70$ km s$^{-1}$ Mpc$^{-1}$ (Freedman et al. 2001), 9 of the sources are ultraluminous. This number does not include the possible central source but does include source number 40.  We estimate that at most two of the nine ULXs are unassociated with the galaxy (Tozzi et al. 2001), and three seem to correspond to globular cluster candidates.  NGC~4697, NGC~1553 and NGC~1399 were found to have 1, 4, and 3 ULXs respectively (Sarazin, Irwin, \& Bregman 2001; Blanton et al. 2001; Angelini et al. 2001).  We find a lower limit on the ratio of the X-ray source luminosity to the optical luminosity of NGC~720 by simply summing the luminosities of the detected sources, excluding the center.  This gives $L_{X_{src}}/L_B$ of $8.1 \times 10^{29}$ ergs s$^{-1}$ $L_{\odot B}^{-1}$.  This number is already larger than the ratios of $7.5 \times 10^{29}$ ergs s$^{-1}$ $L_{\odot B}^{-1}$ and $7.2 \times 10^{29}$ ergs s$^{-1}$ $L_{\odot B}^{-1}$ found for NGC~4697 and NGC~1553 when including an estimate for the unresolved point source emission (Irwin, Sarazin, \& Bregman 2001).  The unresolved emission in NGC~720 is difficult to constrain.  Extrapolating the broken power-law fit to the XLF down to luminosities of $10^{36}$ ergs s$^{-1}$ gives very little increase in the X-ray source luminosity due to the flat low luminosity slope, but using the best-fit single power-law gives almost an order of magnitude increase in the luminosity.  Considering only the ultraluminous sources, the ratio $L_{ULX}/L_B$ for NGC~720 is more than double the ratio for NGC~1553 and a factor of seven higher than for NGC~4697.  If source 40, which could be a background object, is excluded $L_{ULX}/L_B$ for NGC~720 is still 50\% higher than for NGC~1553.  The number of ULXs in NGC~720 is similar to the approximately 10 and 8 detected in the Antennae and NGC~3256 merger systems (Zezas et al. 2002; for $H_0 = 75$ km s$^{-1}$ Mpc$^{-1}$; Lira et al. 2002).  For Antennae, $L_{ULX}/L_B$ is approximately the same as in NGC~720, or 60\% higher if we exclude source 40 from NGC~720.  While individual merger galaxies have been shown to have an overabundance of ULXs, a recent study of 54 galaxies containing ULXs with {\itshape ROSAT\/} found a higher number of these sources per galaxy in ellipticals (Colbert \& Ptak 2002).  For the value of the Hubble constant used in the rest of the paper ($H_0 = 50$ km s$^{-1}$ Mpc$^{-1}$) the number of ULXs in NGC~720 would be 25, and the number in most of the other galaxies discussed would also increase.

Recently, comparison of the XLFs of a few disk and star forming galaxies to bulges and ellipticals has shown that disk/star forming galaxy XLFs seem to have flatter slopes than early-type galaxies implying a higher fraction of ULXs (Kilgard et al. 2002; Zezas \& Fabbiano 2002).  The XLFs in the disk and merger systems studied were fit to single power-laws with slopes between 1.5 and 2.3.  These authors, therefore, argue that these sources may be associated with the young stellar populations present in star forming galaxies.  NGC~720 does not quite conform to this trend.  The best-fit single power-law slope is consistent with those seen in disk galaxies, although a bit steeper than those seen in starbursts.  For the broken power-law fit, the high-end slope of the luminosity function is steep, but the low luminosity slope is 1.4.  With its high break luminosity, the XLF of NGC~720 is flat out to fairly large luminosities.  The large number of ULXs in NGC~720 and its relatively flat XLF may provide evidence against the association of all ULXs with young stars.  As stated above, further evidence for the common occurrence of ULXs in ellipticals is given by the results of Colbert \& Ptak (2002).  Recently, King (2002) suggested that there are two classes of ULXs which both have super-Eddington mass accretion rates: thermal-timescale mass transfer in high-mass X-ray binaries and long-lasting transient outbursts in low-mass X-ray binaries.

\placefigure{fig-5}

\subsection{Spatial Distribution}

We investigated the distribution of the X-ray point sources in NGC~720 in order to compare to the distributions of the X-ray gas, optical light, and globular clusters.  The globular clusters have a similar distribution to the optical light; however, the X-ray gas does not.  The diffuse X-ray emission has an ellipticity of about e $\sim0.15$, and the position angle twists from P.A.$\sim135^{\circ}$ for r$< 50^{\prime\prime}$ to P.A.$\sim110^{\circ}$ at larger radii (Buote et al. 2002).  The stellar light has e $=0.45\pm0.05$ and P.A.$=142^{\circ}\pm3^{\circ}$, and the distribution of GCs follows e $=0.5\pm0.1$ and P.A.$=147^{\circ}\pm10^{\circ}$ (Kissler-Patig, Richtler, \& Hilker 1996; Peletier et al. 1990; Goudfrooij 1994).  Figure 6a shows the angular distribution of X-ray sources.  Angles are measured from north (0$^{\circ}$) to east (90$^{\circ}$).  Also plotted is the predicted distribution of sources for an ellipticity of 0.45 and a position angle of 142$^{\circ}$, following the optical light.  Due to the small number of sources, it is hard to distinguish models, and we find that the X-ray source distribution could be consistent with the optical light, the globular cluster distribution, or the diffuse X-ray emission.

\placefigure{fig-6}

Investigation of the X-ray image, however, reveals that the source distribution appears distinctly non-uniform.  There appear to be arcs of sources as well as large regions with no sources.  Figure 6b shows the angular distribution of sources at distances larger than 90$^{\prime\prime}$ from the galaxy center, while Figure 6c shows the distribution for distances less than 90$^{\prime\prime}$.  At small radii, the distribution is similar to the overall distribution, but at large radii, all of the sources lie in a large arc running from east to south of the galaxy center.  Figure 7 shows a histogram of the source distances from the center of the galaxy.

\placefigure{fig-7}

The distribution of X-ray sources around NGC~720 is quite unusual.  As mentioned before, most of the sources at large radius, which make up the outer arc, are the highly absorbed group A sources.  Approximately a quarter of the sources in NGC~720 are expected to be unassociated with the galaxy, and these sources could be background sources.  However, it would be perhaps more unusual to have such a non-uniform distribution of background sources, and several of these sources are globular cluster candidates.  It is unlikely that the sources would be affected by interaction with a neighboring galaxy.  NGC~720 is fairly isolated; within 20$^{\prime}$, it has two neighbors to the southeast at distances of 19$^{\prime}$ and 15$^{\prime}$ and one to the northwest at 11$^{\prime}$ all of which are much less luminous than NGC~720 (Dressler, Schechter, \& Rose 1986).  The arc pattern of the sources is similar to the ``shells'' seen in some elliptical galaxies at optical wavelengths.  These shell galaxies are thought to be the result of the merger of an elliptical galaxy with a disk galaxy and seem to occur more frequently in isolated ellipticals (Quinn 1984; Colbert, Mulchaey, \& Zabludoff 2001).  NGC~720 is classified as having ``irregular'' isophotes which in conjunction with its strong X-ray emission has been suggested to mean that it has undergone a past merger (Nieto \& Bender 1989).  Perhaps the non-uniform X-ray source distribution is the result of a merger event.  On the other hand, NGC~720 does not show shells at optical wavelengths.  Quinn (1984) estimates that shells will last $\sim1.4h_{70}^{-1}$ Gyrs, and the estimated time since the last major starburst event in NGC~720 from H$\beta$ and [MgFe] absorption line indices is 3.4 Gyrs (Terlevich \& Forbes 2001).

\section{DISCUSSION}

We have detected 42 X-ray point sources within 2.5$^{\prime}$ of the center of the elliptical galaxy NGC~720.  This list includes all sources inside the $D_{25}$ ellipse for this galaxy.  Assuming that they are located at the distance of NGC~720, these sources have luminosities between $4 \times 10^{38}h_{50}^{-2}$ ergs s$^{-1}$ and $10^{40}h_{50}^{-2}$ ergs s$^{-1}$.  Based on the {\itshape Chandra\/} Deep Fields, we estimate that a quarter of these sources are unassociated with the galaxy.  Investigation of the color-color diagram of the sources reveals that most of the sources lie in a diagonal band with negative H31 and H21 ranging from -0.3 to 0.2.  We divide these sources into two groups based on their apparent absorption.  One group of sources is well described by a flat power-law spectrum with a photon index of 1.2 and Galactic absorption.  The other group can be well fit by either a steeper power-law, with a photon index of 2.2 and excess absorption of $2 \times 10^{21}$ cm$^{-2}$, or a flat power-law, with an index of 1.1 plus a thermal blackbody component with a temperature of 0.38$^{+0.07}_{-0.06}$ keV.  There is also a group of sources with positive hardness ratios, indicating that most of their counts are at high energies.  The combined spectrum of this group is well fit by a power-law with an index of about 2 and high absorption of $7 \times 10^{21}$ cm$^{-2}$ which cannot be accounted for by an additional thermal component.  These highly absorbed sources are generally located at large distances from the galaxy center and could be mostly background sources.  However, four of the these sources seem to correspond to globular cluster candidates.  The spectral fits are summarized in Table 3.  We also detect a source near the center of NGC~720 which could be a central low-luminosity AGN.  This source has large hardness ratios and a luminosity of $2.5 \times 10^{39}h_{50}^{-2}$ ergs s$^{-1}$.

The overall spatial distribution of X-ray sources in NGC~720 is consistent with the galaxy ellipticity and position angle from either the diffuse X-ray emission or optical light.  However, the sources do not appear to be uniformly distributed.  For example, at large radii all of the sources lie in a large arc running from east to south of the galaxy center.  NGC~720 is fairly isolated and it is unlikely that the source distribution could be explained by interaction with another galaxy.  It is possible that this feature could be due to a background structure, but it seems more likely that such a non-uniform distribution would be associated with the galaxy.  Most of the sources in this arc belong to the group of sources with large hardness ratios and high absorption, and the feature may be the result of a past merger and starburst event.

With {\itshape Chandra\/}, the luminosity functions of X-ray sources in several galaxies have been studied.  In two other early-type galaxies the XLFs have been well fit by a broken power-law with a break luminosity of 3-4 $\times 10^{38}h_{50}^{-2}$ ergs s$^{-1}$ and a high luminosity slope of 2.7-2.8 (Sarazin et al. 2000; Blanton et al. 2001).  In contrast, the disk and merger galaxies studied so far have XLFs which can be fit by a single power-law with a flatter slope (Kilgard et al. 2002; Zezas \& Fabbiano 2002).  We do not detect sources to low enough luminosities in NGC~720 to be able to compare to the break seen in NGC~4697 and NGC~1553; however, we see possible evidence of a break at a much higher luminosity of $2 \times 10^{39}h_{50}^{-2}$ ergs s$^{-1}$.  This break is too high to be explained as a transition between neutron star and black hole binaries as suggested by Sarazin et al. (2000).  Alternatively, it has been suggested that a break in the XLF could be produced by a decaying starburst component from binaries formed in a past merger (Kilgard et al. 2002; Wu 2001).  If this is the case, NGC~720 may have undergone a more recent merger than either NGC~4697 or NGC~1553.  A broken power-law is a somewhat better fit to the NGC~720 XLF than a single power-law, with a low luminosity slope of 1.4 and a high luminosity slope of 5.  With its high break luminosity, the XLF of NGC~720 is flat out to fairly large luminosities.  When fit to a single power-law, the slope of the XLF in NGC~720 is possibly flatter than the other early-type galaxies and possibly steeper than the disk and merger galaxies, but could agree with either within the errors.  This flatness of the XLF also agrees with the possiblity of an evolved starburst from a past merger.  The ``shell-like'' distribution of X-ray sources in NGC~720 could be the result of a past merger.  However, the sources in NGC~720 would have to be associated with relatively low mass stars.  As mentioned before, the time since the last major starburst event in NGC~720 is estimated to be 3.4 Gyrs (Terlevich \& Forbes 2001).  In addition, the formation time of shells in a merger is 0.1-1 Gyrs (Quinn 1984).  We would, therefore, only expect to find main-sequence stars of 1-2 M$_{\odot}$ or less.

Ultraluminous sources (L$_{x} \geq 10^{39}$ ergs s$^{-1}$) have now been seen in galaxies of many morphological types, and the nature of these sources is unknown.  They could be associated with 10-100 M$_{\odot}$ black holes binaries (e.g. Colbert \& Mushotzky 1999) or beamed emission from stellar mass black hole or neutron star binaries (King et al. 2001; King 2002).  Comparison of the X-ray luminosity functions of early-type and star forming galaxies suggests that perhaps there is a larger fraction of these sources in star forming galaxies and that they could be associated with young stellar populations (Kilgard et al. 2002; Zezas \& Fabbiano 2002).  NGC~720 may provide evidence against this suggestion.  Fitting to a single power-law, its luminosity function is nearly as flat as those seen in disk and merger systems, and fitting to a broken power-law, the XLF is extremely flat out to a luminosity of $2 \times 10^{39}h_{50}^{-2}$ ergs s$^{-1}$ before dropping off.  It also contains a large number of ultraluminous sources.  We find 9 ULXs in NGC~720 for $H_0 = 70$ km s$^{-1}$ Mpc$^{-1}$ of which at most two are expected to be unassociated with the galaxy.  This number is similar to the 10 and 8 detected in the Antennae and NGC~3256 merger systems (Zezas et al. 2002; for $H_0 = 75$ km s$^{-1}$ Mpc$^{-1}$; Lira et al. 2002) and more than has been previously seen in an early-type galaxy.  

We investigated the possibility that we could be overestimating the number of ULXs in NGC~720.  A 20\% error in the distance would reduce the number of ULXs to 5.  We also calculated source luminosities using different spectral models including absorption fixed at Galactic and the other fits listed in Table 3.  Dividing the sources into groups A, B, and C and varying the spectral models gave a minimum of 6 ULXs.  Using the composite spectrum of all of the sources and varying the model gave a minimum of 4 ULXs.  A reduction in the number of ultraluminous sources to 4 would be in good agreement with the number seen in NGC~1553.  The detection of ULXs in several early-type galaxies implies that not all of these systems are related to young, massive stars.  In a study of 54 galaxies containing ULXs with {\itshape ROSAT\/}, Colbert \& Ptak (2002) found a higher number of these sources per galaxy in ellipticals, and King (2002) suggests that there may be two classes of ULXs with one class associated with HMXBs and the other with LMXBs.  However, the unusual source distribution may indicate that NGC~720 has undergone a merger, provding a possible connection between these sources and those seen in the Antennae.  Again, the sources in NGC~720 would have to be associated with lower mass stars of at most 1-2 M$_{\odot}$.

So far only a handful of galaxy source populations have been studied.  As a larger sample of galaxies becomes available with {\itshape Chandra\/}, the properties of ULXs and their relationship to galaxy morphology as well as the location and cause of the break in the XLFs of early-type galaxies can be studied with more certainty.

After this paper was submitted, evidence was published for the existence of intermediate mass black holes at the center of two globular clusters, G1 and M15 (Gebhardt, Rich, \& Ho 2002; Gerssen et al. 2002a; Gerssen et al. 2002b; van der Marel et al. 2002).  This discovery is relevant to the discussion of the nature of ULXs.  It is interesting to note that in NGC~720 three of the ULXs are close to globular cluster candidates.

\acknowledgments
We would like to thank MIT HETG/CXC group, particularly Herman Marshall for all his help in the analysis.  We would also like to thank Markus Kissler-Patig for supplying the GC list, and Paul Schechter for several helpful discussions.  This work was funded by NASA contracts NAS8-38252, NAS8-37716, and NAS8-38249.  TEJ would like to acknowledge the support of a NSF fellowship.

\clearpage
\begin{deluxetable}{lc}
\tabletypesize{\scriptsize}
\tablecaption{Basic Properties \label{tbl-1}}
\tablewidth{0pt}
\tablecolumns{2}
\tablehead{
\colhead{Parameter} & \colhead{Value}
}
\startdata
$\alpha, \delta$ (J2000) &1:53:00.46, -13:44:18.4\\
Hubble Type &E5\\
$m^0_B$ &11.13\\
$D_{25}$ &4$^{\prime}$41$^{\prime\prime}$\\
T$_X$ &0.6 keV\\
L$_X$ (0.7-2 keV) &$5.2\times10^{40}h_{50}^{-2}$ ergs/s\\
\enddata



\tablecomments{The position listed is the position of the optical center as given in Smith et al. (2000).  The X-ray temperature and luminosity are from the {\itshape Chandra\/} data and include the diffuse emission only (Buote et al. 2002).  This temperature agrees well with previous fits to the {\itshape ROSAT\/} and {\itshape ASCA\/} spectra.  The luminosity is calculated from the spectral analysis in Buote et al. (2002) for a radius of 142$^{\prime\prime}$.  All other parameters are taken from the RC3 (de Vaucouleurs et al. 1991)}

\end{deluxetable}

\clearpage
\begin{deluxetable}{lccccccc}
\tabletypesize{\scriptsize}
\tablecaption{X-ray Point Sources \label{tbl-2}}
\tablewidth{0pt}
\tablecolumns{8}
\tablehead{
\colhead{Source} & \colhead{R.A.}   & \colhead{Dec.}   &
\colhead{d} & \colhead{Count Rate}   &
\colhead{Significance} & \colhead{Luminosity (0.3-10 keV)} & \colhead{Notes} \\ & \colhead{(J2000)} & \colhead{(J2000)} & \colhead{(arcsec)} & \colhead{(10$^{-4}$ s$^{-1}$)} &\colhead{($\sigma$)} & \colhead{(10$^{38}h_{50}^{-2}$ ergs s$^{-1}$)}
}
\startdata
1 &1:53:00.42 &-13:44:19.7 &1.4 &10.1$\pm$3.0 &5 &25.25 &A\\
2 &1:53:00.29 &-13:44:13.5 &5.2 &8.5$\pm$2.8 &5 &11.05 &C\\
3 &1:53:00.08 &-13:44:11.7 &8.7 &7.0$\pm$2.2 &5 &7.70 &B\\
4 &1:53:01.14 &-13:44:19.7 &10.0 &15.8$\pm$3.1 &$>$8 &17.38 &B\\
5 &1:53:00.84 &-13:44:06.8 &12.8 &16.1$\pm$2.9 &$>$8 &20.93 &C\\
6 &1:53:00.80 &-13:44:00.0 &19.0 &6.7$\pm$2.1 &8 &16.75 &A\\
7 &1:53:00.29 &-13:43:55.7 &23 &12.1$\pm$2.4 &$>$8 &15.73 &GC, C\\
8 &1:52:59.32 &-13:44:02.4 &23 &4.3$\pm$1.6 &5 &4.73 &B\\
9 &1:52:59.41 &-13:43:56.9 &26 &27.9$\pm$3.2 &$>$8 &36.27 &GC, C\\
10 &1:53:01.98 &-13:44:35.0 &28 &8.9$\pm$2.1 &8 &11.57 &C\\
11 &1:53:01.22 &-13:44:47.3 &31 &8.4$\pm$2.1 &8 &9.24  &B\\
12 &1:53:00.84 &-13:44:49.2 &31 &9.6$\pm$2.1 &8 &10.56 &B\\
13 &1:53:02.49 &-13:44:34.4 &34 &4.3$\pm$1.7 &4 &5.59 &C\\
14 &1:53:01.85 &-13:44:45.5 &34 &3.9$\pm$1.5 &5 &4.29 &GC, B\\
15 &1:52:59.32 &-13:44:48.6 &34 &14.0$\pm$2.5 &$>$8 &18.20 &GC, C\\
16 &1:53:00.24 &-13:43:43.4 &35 &3.9$\pm$1.5 &8 &4.29 &GC, B\\
17 &1:53:02.99 &-13:44:25.8 &38 &3.9$\pm$1.5 &8 &5.07 &C\\
18 &1:52:58.94 &-13:44:49.2 &38 &4.5$\pm$1.8 &5 &4.95 &GC, B\\
19 &1:52:58.39 &-13:43:42.8 &47 &6.0$\pm$1.8 &8 &7.8 &C\\
20 &1:52:58.43 &-13:44:56.6 &48 &3.8$\pm$1.7 &4 &4.18 &B\\
21 &1:53:02.44 &-13:44:57.8 &49 &3.6$\pm$1.6 &4 &3.96 &B\\
22 &1:52:56.45 &-13:43:47.7 &66 &15.4$\pm$2.8 &$>$8 &38.50 &GC, A\\
23 &1:52:56.03 &-13:44:39.3 &68 &8.7$\pm$2.1 &8 &21.75 &A\\
24 &1:52:56.41 &-13:43:40.3 &70 &9.2$\pm$2.0 &8 &10.12 &GC, B\\
25 &1:53:02.66 &-13:43:15.1 &71 &3.5$\pm$1.5 &5 &8.75 &A\\
26 &1:52:56.66 &-13:43:31.1 &73 &8.1$\pm$2.2 &8 &8.91 &B\\
27 &1:52:55.78 &-13:43:50.8 &73 &19.0$\pm$3.0 &$>$8 &20.90 &B\\
28 &1:52:57.67 &-13:45:20.5 &74 &4.7$\pm$1.9 &5 &11.75 &A\\
29 &1:53:01.98 &-13:42:49.9 &91 &14.1$\pm$2.4 &$>$8 &35.25 &A\\
30 &1:53:06.75 &-13:43:58.8 &94 &4.7$\pm$1.8 &5 &6.11 &C\\
31 &1:53:06.66 &-13:45:01.5 &100 &3.9$\pm$1.6 &5 &9.75 &GC, A\\
32 &1:53:06.71 &-13:43:35.4 &101 &5.0$\pm$1.8 &8 &12.50 &A\\
33 &1:53:07.13 &-13:43:23.7 &111 &5.0$\pm$1.6 &8 &12.50 &A\\
34 &1:53:08.56 &-13:44:10.4 &118 &4.0$\pm$1.8 &4 &10.00 &A\\
35 &1:53:06.75 &-13:45:34.7 &119 &13.0$\pm$2.5 &$>$8 &32.50 &A\\
36 &1:53:06.44 &-13:45:40.9 &120 &20.7$\pm$2.8 &$>$8 &26.91 &C\\
37 &1:53:05.90 &-13:45:54.4 &124 &2.5$\pm$1.5 &3 &6.25 &A\\
38 &1:53:08.61 &-13:44:56.6 &125 &14.2$\pm$2.4 &$>$8 &15.62 &GC, B\\
39 &1:53:03.88 &-13:46:13.4 &125 &9.5$\pm$2.2 &8 &10.45 &B\\
40 &1:53:01.77 &-13:46:30.7 &134 &46.1$\pm$4.1 &$>$8 &109.8 &GC, A\\
41 &1:53:04.51 &-13:46:22.7 &138 &6.1$\pm$2.2 &5 &15.25 &GC, A\\
42 &1:53:10.13 &-13:43:28.6 &149 &10.5$\pm$2.2 &8 &13.65 &C\\
\enddata



\tablecomments{In column 4, d is the radial distance from the optical center of the galaxy.  In the last column, ``GC'' indicates that the source is located within 2$^{\prime\prime}$ of a globular cluster candidate.  ``A'', ``B'', and ``C'' refer to source group.  Groups were selected based on hardness ratio as described in Section 4.1.  Luminosities are calculated for a distance to NGC~720 of $35^{+7}_{-6} h_{50}^{-1}$ Mpc.}

\end{deluxetable}

\clearpage
\begin{deluxetable}{lcccccc}
\tabletypesize{\scriptsize}
\tablecaption{Spectral Fits \label{tbl-3}}
\tablewidth{0pt}
\tablecolumns{7}
\tablehead{
\colhead{Sources} & \colhead{Model}   & \colhead{N$_H$}  & \colhead{$\Gamma$ or kT} & \colhead{Second Model} & \colhead{kT} & \colhead{$\chi^2_{\nu}$/DOF}\\ 
& & \colhead{($10^{21}$ cm$^{-2}$)} & \colhead{(keV)} & & \colhead{(keV)} &
}
\startdata
All Sources &power &1.1$^{+0.4}_{-0.3}$ &1.5$\pm0.1$ &\nodata &\nodata &0.88/85\\
All Sources &power &(0.155) &1.2$\pm0.2$ &bbody &0.62$^{+0.15}_{-0.11}$ &0.86/84\\
All Sources &bremss &0.8$\pm0.3$ &11$^{+13}_{-4}$ &\nodata &\nodata &0.86/85\\
A &power &7$^{+3}_{-2}$ &1.9$^{+0.5}_{-0.3}$ &\nodata &\nodata &1.00/23\\
A &bremss &6$\pm2$ &5$^{+6}_{-2}$ &\nodata &\nodata &0.98/23\\
B &power &(0.155) &1.2$\pm0.2$ &\nodata &\nodata &0.60/34\\
C &power &2.0$\pm0.6$ &2.2$\pm0.2$ &\nodata &\nodata &0.83/36\\
C &power &(0.155) &1.1$\pm0.3$ &bbody &0.38$^{+0.07}_{-0.06}$ &0.85/35\\
C &bremss &1.1$^{+0.5}_{-0.3}$ &2.9$^{+1.4}_{-0.8}$ &\nodata &\nodata &0.83/36\\
Source 40 &power &4$\pm2$ &1.3$\pm0.3$ &\nodata &\nodata &1.05/8\\
Source 40 &bbody &(0.155) &0.93$^{+0.16}_{-0.13}$ &\nodata &\nodata &0.80/9\\
\enddata



\tablecomments{Fixed parameters are enclosed in parentheses.}

\end{deluxetable}

\clearpage
\begin{figure}
\figurenum{1}
\figcaption{({\itshape Left\/}) Smoothed {\itshape Chandra\/} image of NGC~720 in the 0.3-10 keV energy band.  The image was smoothed with the CIAO program {\itshape csmooth\/}.  ({\itshape Right\/}) The same image overlaid with the optical contours from the DSS image.  Contours are linearly spaced. \label{fig-1}}
\end{figure}

\begin{figure}
\figurenum{2}
\plotone{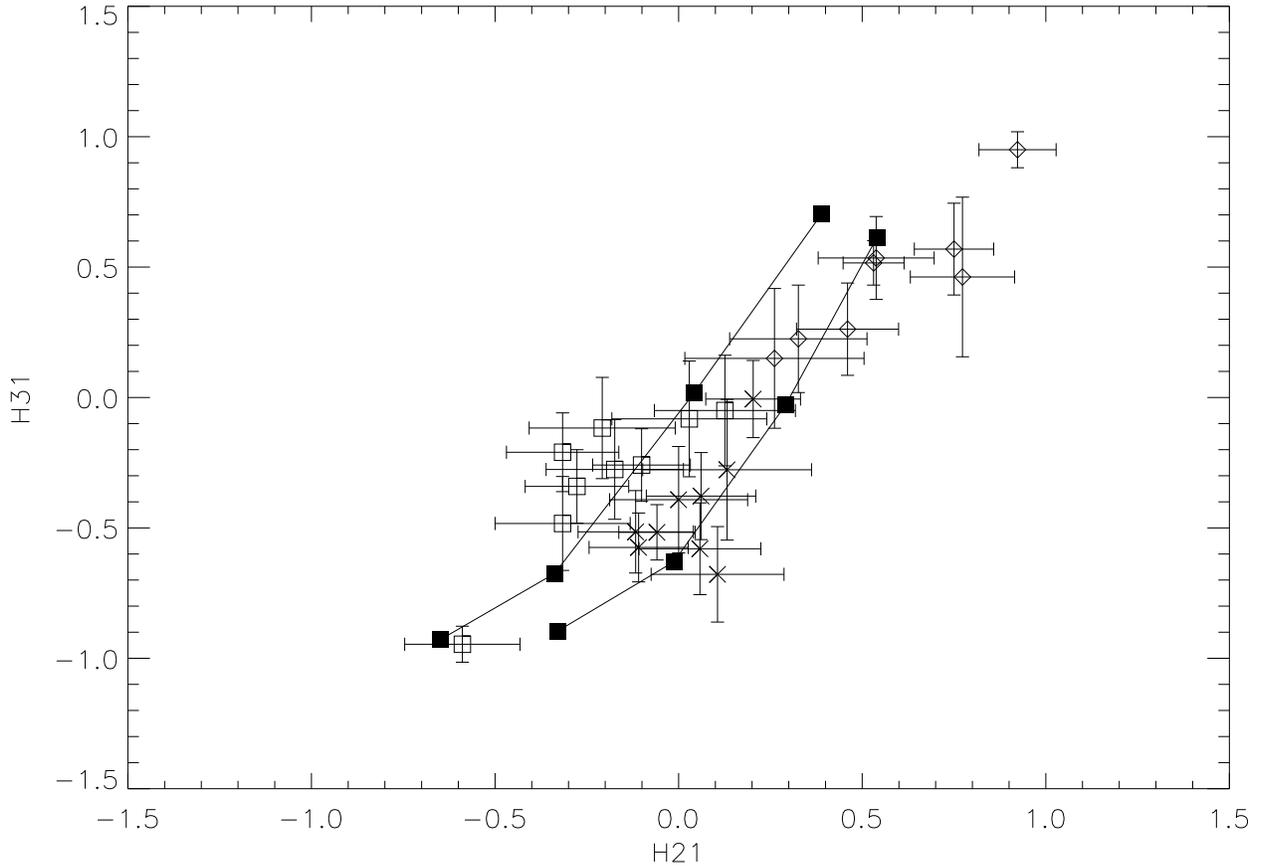}
\figcaption{Color-color diagram of sources with more than 20 counts in the 0.3-10.0 keV band.  The hardness ratios are defined as H21 = (M-S)/(M+S) and H31= (H-S)/(H+S).  Error bars give 1-sigma errors in the ratios.  The upper line shows predicted ratios for a power-law with a hydrogen column density equal to the Galactic value of $1.55 \times 10^{20}$ cm$^{-2}$; from top to bottom, the plotted points are for indices of 0, 1, 2, and 3.  The lower line gives the ratios for a column of $3.0 \times 10^{21}$ cm$^{-2}$ and indices of 1, 2, 3, and 4. Group A sources are plotted with open diamonds, group B sources are plotted with open squares, and group C sources are plotted with Xs. \label{fig-2}}
\end{figure}

 \begin{figure}
\figurenum{3}
\epsscale{0.5}
\plotone{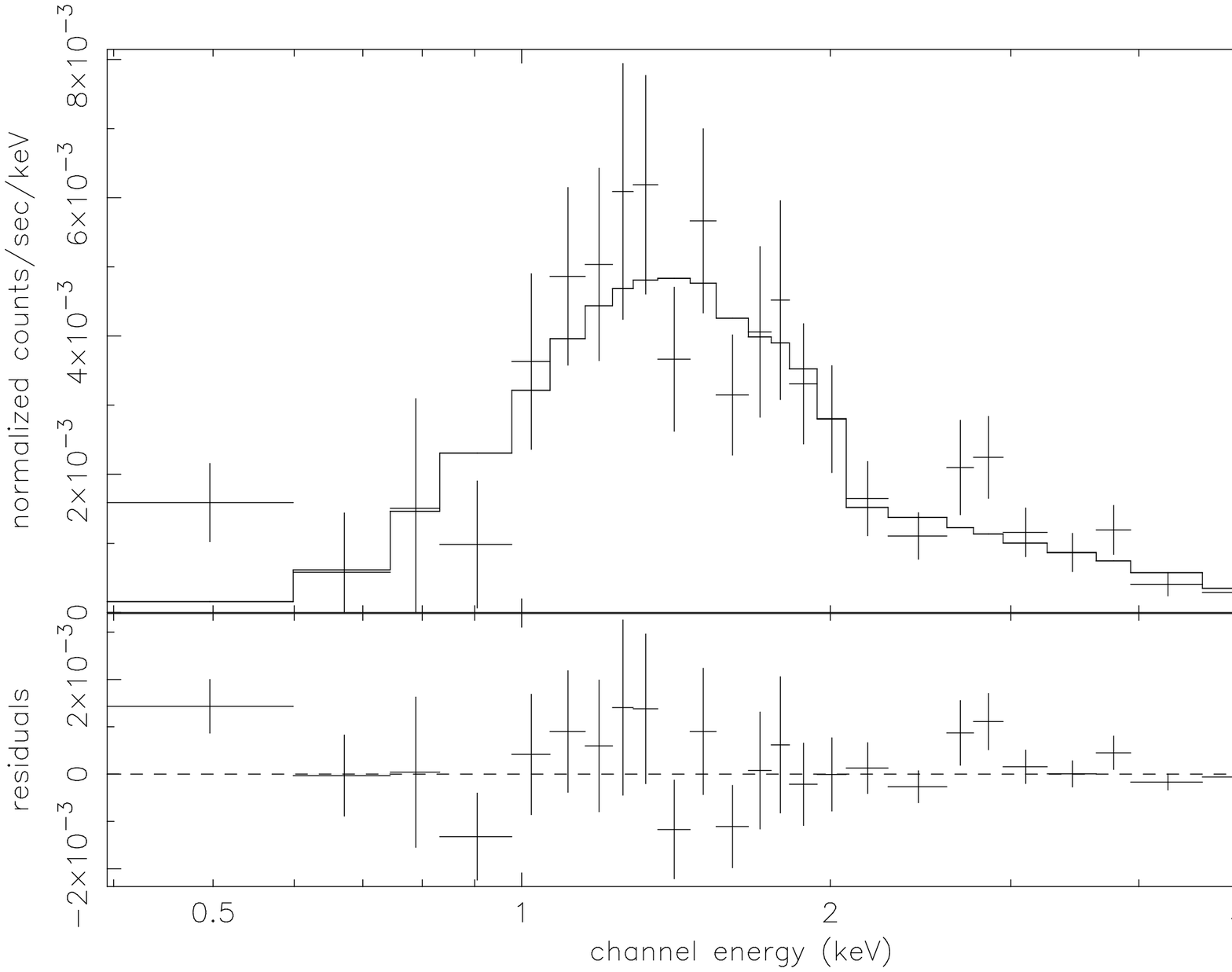}
\plotone{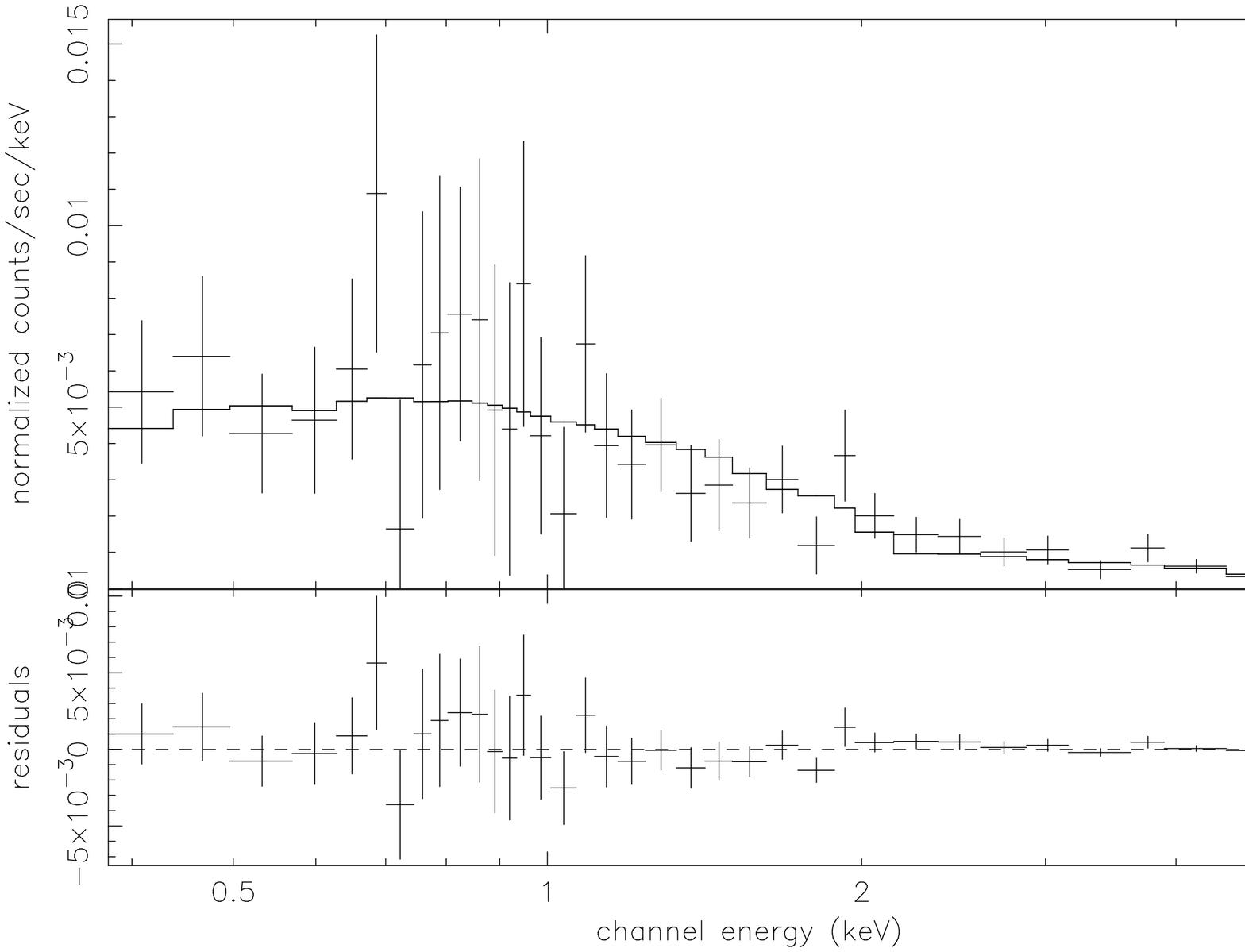}
\plotone{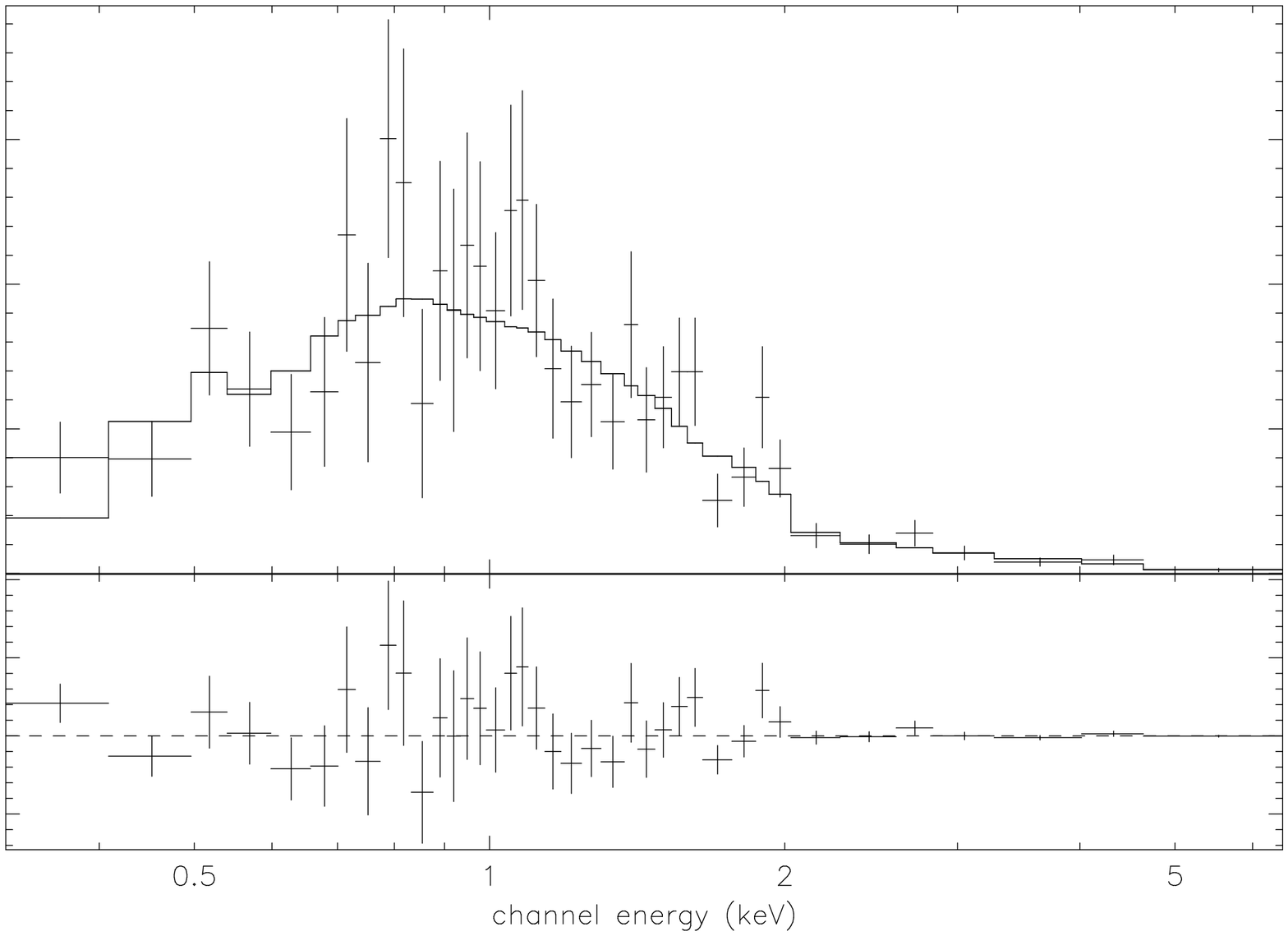}
\figcaption{Composite spectra of A (top), B (middle), and C (bottom) sources.  For each plot, the top panel shows the spectrum and best-fit power-law model, and the bottom panel shows the residuals.  For the B spectrum the N$_{H}$ was fixed at the galactic value, and for the other two spectra it was allowed to vary. \label{fig-3}}
\end{figure}

\begin{figure}
\figurenum{4}
\epsscale{0.8}
\plotone{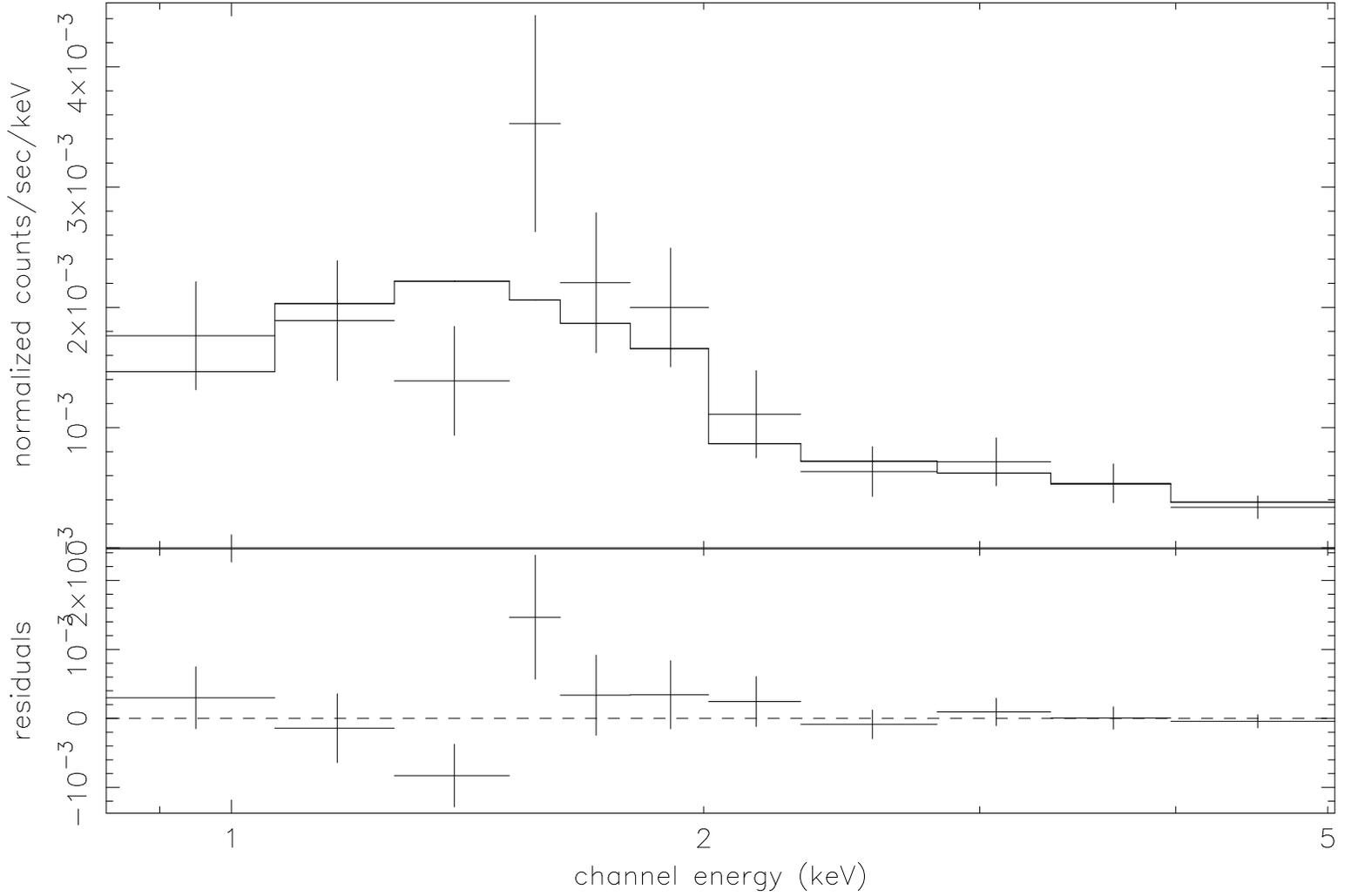}
\figcaption{Spectrum of the brightest source, source 40.  The top panel shows the best-fit power-law model, and the bottom panel shows the residuals. \label{fig-4}}
\end{figure}

\begin{figure}
\figurenum{5}
\epsscale{1.0}
\plotone{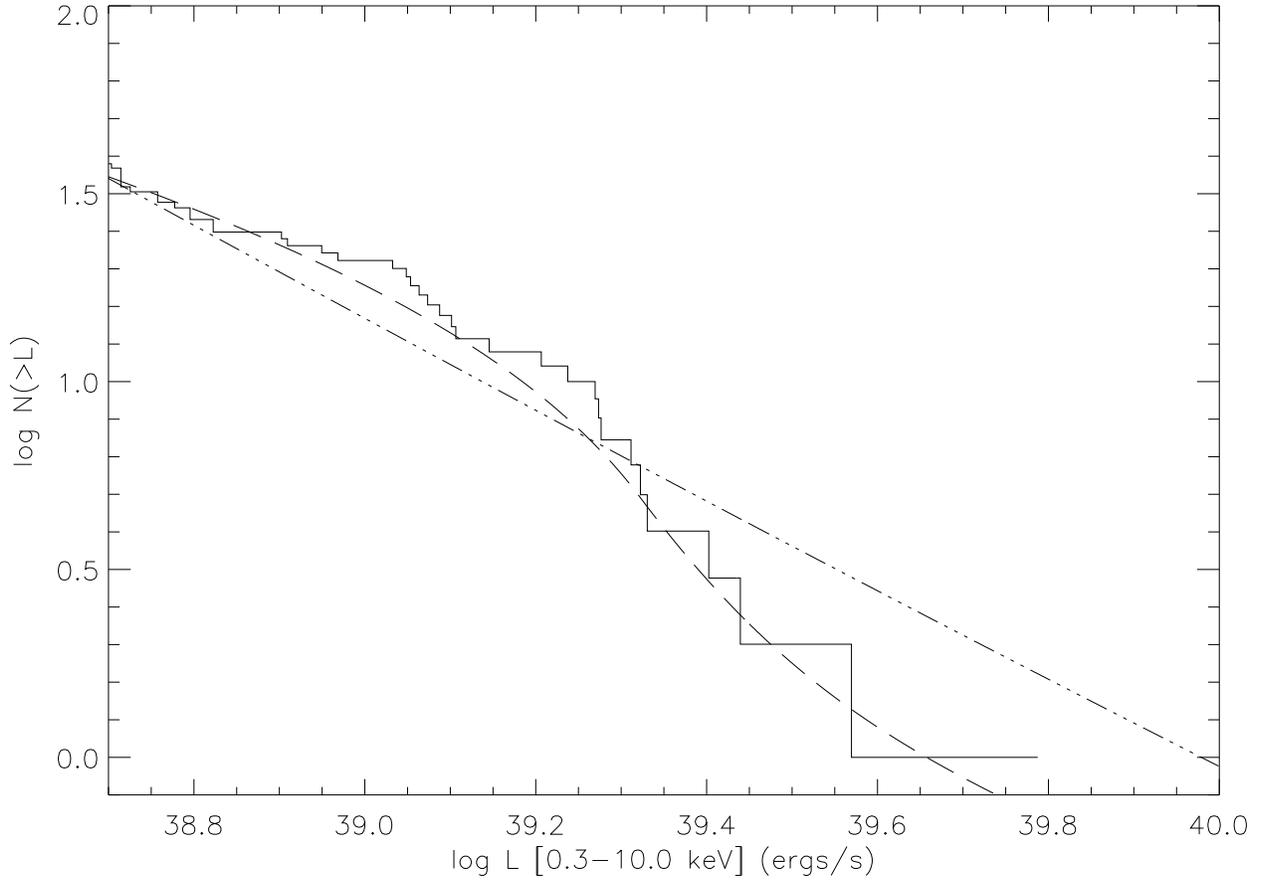}
\figcaption{Luminosity function of the sources.  The dashed curve shows the best-fitting broken power-law to the differential distribution.  This model includes a term which models the contribution from background sources.  The dot-dashed curve shows the best-fit single power-law plus background. \label{fig-5}}
\end{figure}

\begin{figure}
\figurenum{6}
\epsscale{1.0}
\plotone{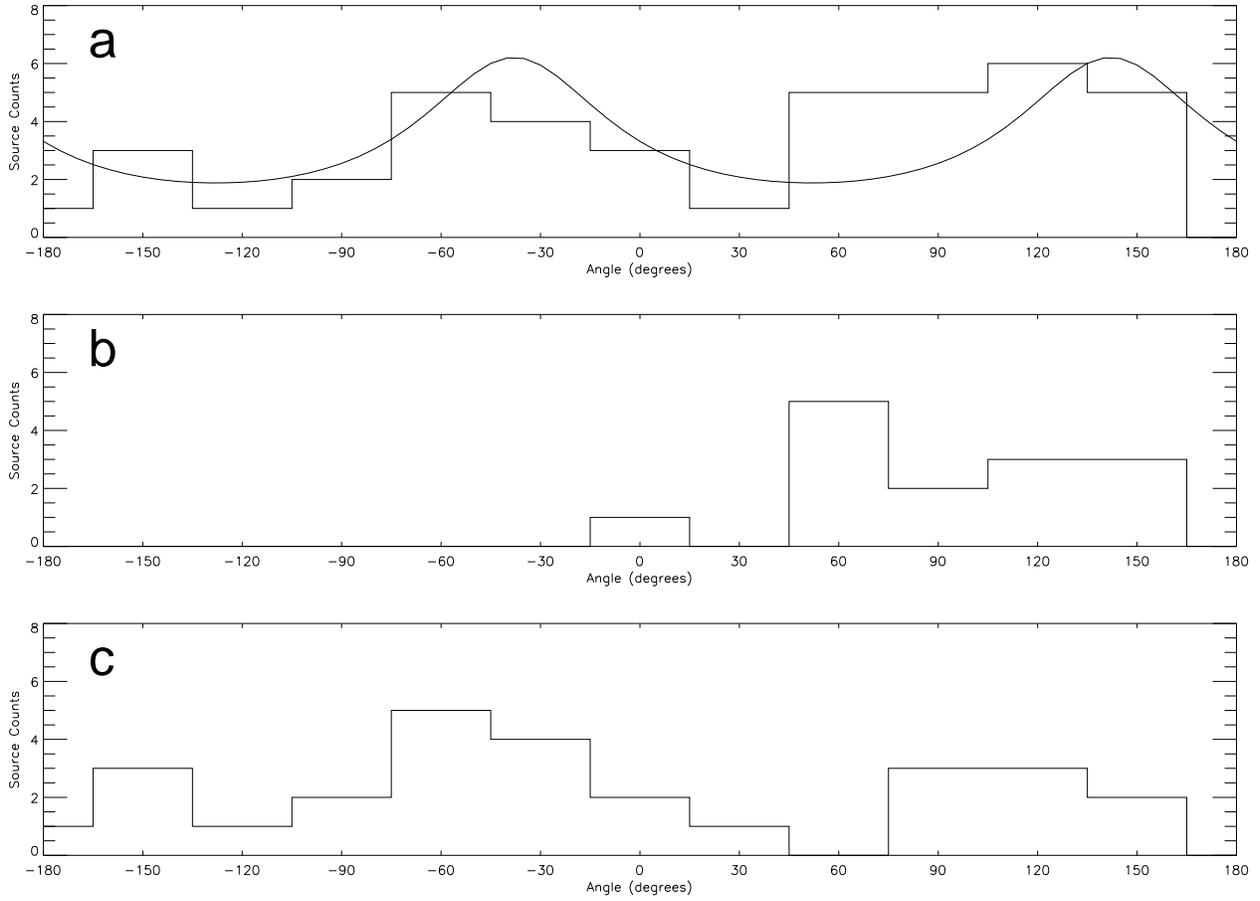}
\figcaption{Histogram of the angular distribution of X-ray sources with a binning of 30$^{\circ}$.  Angles are measured from north (0$^{\circ}$) to east (90$^{\circ}$).  The distribution of all sources is shown in (a) along with a model for the predicted distribution based on the distribution of the optical light.  Sources at r $> 90^{\prime\prime}$ are shown in (b), and sources at r $< 90^{\prime\prime}$ in (c). \label{fig-6}}
\end{figure}

\begin{figure}
\figurenum{7}
\plotone{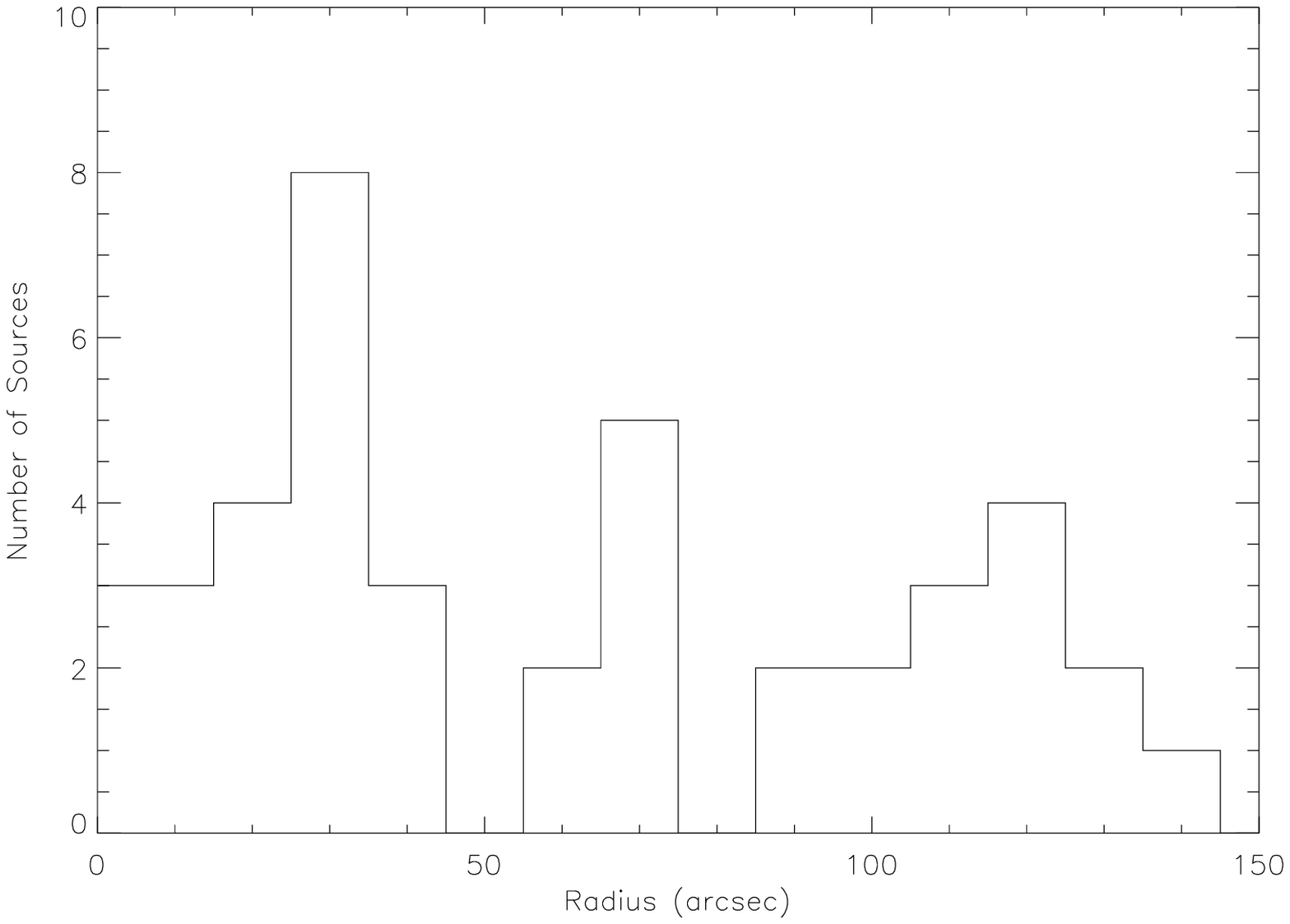}
\figcaption{Histogram of source distances from the optical center of NGC~720 with 10$^{\prime\prime}$ bins. \label{fig-7}}
\end{figure}


\begin{thebibliography}{}

\bibitem[]{152} Angelini, L., Loewenstein, M., \& Mushotzky, R. F. 2001, ApJ, 557, L35

\bibitem[]{154} Baldi, A., Molendi, S., Comastri, A., Fiore, F., Matt, G., \& Vignali, C. 2001, ApJ, 564, 190

\bibitem[]{156} Blanton, E. L., Sarazin, C. L., Irwin, J. A. 2001, ApJ, 552, 106

\bibitem[]{158} Blakeslee, J. P., Lucey, J. R., Barris, B. J., Hudson, M. J., \& Tonry, J. L. 2001, MNRAS, 327, 1004

\bibitem[]{160} Blakeslee, J. P., Lucey, J. R., Tonry, J. L., Hudson, M. J., Narayanan, V. K., \& Barris, B. J. 2002, MNRAS, 330, 443

\bibitem[]{162} Brandt, W. N. et al. 2001, AJ, 122, 2810

\bibitem[]{164} Buote, D. A., \& Canizares, C. R. 1994, ApJ, 427, 86

\bibitem[]{166} Buote, D. A., \& Canizares, C. R. 1996, ApJ, 468, 184

\bibitem[]{168} Buote, D. A., \& Canizares, C. R. 1997, ApJ, 474, 650

\bibitem[]{170} Buote, D. A., Jeltema, T. E., Canizares, C. R., \& Garmire, G. P. 2002, ApJ, 577, 183

\bibitem[]{172} Campana, S., Moretti, A., Lazzati, D., \& Tagliaferri, G. 2001, ApJ, 560, L19

\bibitem[]{174} Chapelan, S. 2001, PhD Thesis, University de Provence

\bibitem[]{176} Colbert, J. W., Mulchaey, J. S., \& Zabludoff, A. I. 2001, AJ, 121, 808

\bibitem[]{183} Colbert, J. W., \& Mushotzky, R. F. 1999, ApJ, 519, 89

\bibitem[]{185} Colbert, E. J. M., \& Ptak, A. F. 2002, ApJS, accepted, astro-ph/0204002

\bibitem[]{178} Dickey, J. M., \& Lockman, F. J. 1990, ARA\&A, 28, 215

\bibitem[]{180} Dressler, A., Schechter, P. L., \& Rose, J. A. 1986, AJ, 91, 105

\bibitem[]{182} Fabbiano, G. 1989, Ann. Rev. Ast. Ap., 27, 87

\bibitem[]{184} Fabbiano, G., Kim, D.-W., \& Trinchieri, G. 1994, ApJ, 429, 94

\bibitem[]{186} Fabbiano, G., Zezas, A., \& Murray, S. S. 2001, ApJ, 554, 1035

\bibitem[]{190} Forman, W., Jones, C., \& Tucker, W. C. 1985, ApJ, 293, 102

\bibitem[]{192} Freedman et al. 2001, ApJ, 553, 47

\bibitem[]{} Gebhardt, K., Rich, R. M., \& Ho, L. C. 2002, ApJ, 578, L41

\bibitem[]{201} Gehrels, N. 1986, ApJ, 303, 336

\bibitem[]{} Gerssen, J., van der Marel, R. P., Gebhardt, K., Guhathakurta, P., Peterson, R. C., \& Pryor, C. 2002a, AJ, in press, astro-ph/0209315

\bibitem[]{} Gerssen, J., van der Marel, R. P., Gebhardt, K., Guhathakurta, P., Peterson, R. C., \& Pryor, C. 2002b, AJ, submitted, astro-ph/0210158

\bibitem[]{194} Goudfrooij, P., Hansen, L., Jorgensen, H. E., Norgaard-Nielsen, K. U., de Jong, T., \& van den Hoek, L. B. 1994, A\&AS 104, 179

\bibitem[]{196} Ho, L. C., et al. 2001, ApJ, 549, L51

\bibitem[]{198} Irwin, J. A., Sarazin, C. L., \& Bregman, J. N. 2002, ApJ, 570, 152

\bibitem[]{} Irwin, J. A., \& Sarazin, C. L. 1998, ApJ, 499, 650

\bibitem[]{209} Kilgard, R. E., Kaaret, P., Krauss, M. I., Prestwich, A. H., Raley, M. T., \& Zezas, A. 2002, ApJ, 573, 138

\bibitem[]{} Kim, D.-W., Fabbiano, G., \& Trinchieri, G. 1992, ApJ, 393, 134

\bibitem[]{200} King, A. R., Davies, M. B., Ward, M. J., Fabbiano, G., \& Elvis, M. 2001, ApJ, 552, L109

\bibitem[]{213} King, A. R. 2002, MNRAS Letters, accepted, astro-ph/0206117

\bibitem[]{202} Kissler-Patig, M., Richtler, T., \& Hilker, M. 1996, A\&A, 308, 704

\bibitem[]{217} Kormendy, J., \& Gebhardt, K. 2001, in AIP Conf. Proc. 586, 20th Texas Symposium on relativistic astrophysics, ed. H. Martel \& J.C. Wheeler (Melville, NY: AIP), 363

\bibitem[]{219} Lira, P., Ward, M., Zezas, A., Alonso-Herrero, A., \& Ueno, S. 2002, MNRAS, 330, 259

\bibitem[]{} Loewenstein, M., Mushotzky, R. F., Angelini, L., Arnaud, K. A., \& Quataert, E. 2001, ApJ, 555, L21

\bibitem[]{} Matsumoto, H., Koyama, K., Awaki, H., Tsuru, T., Loewenstein, M., \& Matsushita, K. 1997, 482, 133

\bibitem[]{204} Mushotzky, R. F., Cowie, L. L., Barger, A. J., \& Arnaud, K. A. 2000, Nature, 404, 459

\bibitem[]{206} Nieto, J. -L., Bender, R. 1989, A\&A 215, 266

\bibitem[]{208} Peletier, R. F., Davies, R. L., Illingworth, G. D., Davis, L. E., \& Cawson, M. 1990, AJ, 100, 1091

\bibitem[]{212} Quinn, P. J. 1984, ApJ, 279, 596

\bibitem[]{214} Sarazin, C. L., Irwin, J. A., \& Bregman, J. N. 2000, ApJ, 544, L101

\bibitem[]{216} Sarazin, C. L., Irwin, J. A., \& Bregman, J. N. 2001, ApJ, 556, 533

\bibitem[]{218} Smith, R. J., Lucey, J. R., Hudson, M. J., Schlegel, D. J., Davies, R. L. 2000, MNRAS, 313, 469

\bibitem[]{} Terashima, Y., \& Wilson, A. S. 2002, ApJ, accepted, astro-ph/0209607

\bibitem[]{220} Terlevich, A. I., \& Forbes, D. A. 2001, MNRAS, 330, 547

\bibitem[]{222} Tonry, J. L., Dressler, A., Blakeslee, J. P., Ajhar, E. A., Fletcher A. B., Luppino, G. A., Metzger, M. R., \& Moore, C. B. 2001, ApJ, 546, 681

\bibitem[]{224} Townsley, L. K., Broos, P. S., Garmire, G. P., \& Nousek, J. A. 2000, ApJ, 534, L139

\bibitem[]{226} Tozzi, P. et al. 2001, ApJ, 562, 42

\bibitem[]{243} van der Marel, R. P. 1999, AJ, 117, 744

\bibitem[]{} van der Marel, R. P., Gerssen, J., Guhathakurta, P., Peterson, R. C., \& Gebhardt, K. 2002, AJ, in press, astro-ph/0209314

\bibitem[]{245} Vaucouleurs, de G., Vaucouleurs, de A., Corwin, H. G., Buta, R. J., Paturel, G., Fouque, P. 1991, Third Reference Catalog of Bright Galaxies, Springer, New York

\bibitem[]{228} Wu, K. 2001, accepted for publication in Publ. Astron. Soc. Aus., astro-ph/0103157

\bibitem[]{249} Zezas, A. Fabbiano, G., Rots, A. H., \& Murray, S. S. 2002, ApJS, 142, 239

\bibitem[]{251} Zezas, A. \& Fabbiano, G. 2002, ApJ, submitted, astro-ph/0203176

\end{thebibliography}
\end{document}